\documentclass[12pt,a4j]{article}
\usepackage[dvips]{graphicx}
\usepackage{enumerate}
\usepackage{amsmath}

\begin{document}
\baselineskip=7mm
\centerline{\bf  A novel multi-component generalization of the short pulse equation  }\par
\centerline{\bf  and its multisoliton solutions} \par
\bigskip
\centerline{Yoshimasa Matsuno\footnote{{\it E-mail address}: matsuno@yamaguchi-u.ac.jp}}\par

\centerline{\it Division of Applied Mathematical Science, Graduate School of Science and Engineering} \par
\centerline{\it Yamaguchi University, Ube, Yamaguchi 755-8611, Japan} \par
\bigskip
\bigskip
We propose a novel multi-component system of nonlinear equations that generalizes the short pulse (SP) equation
describing the propagation of ultra-short pulses in optical fibers.
By means of the bilinear formalism combined with a hodograph transformation, we obtain 
its multi-soliton solutions in the form of a parametric representation. 
Notably, unlike the determinantal solutions of the SP equation, the proposed system is found to exhibit  solutions expressed in terms of pfaffians.
The proof of the solutions is performed within the framework of  an elementary theory of determinants.
The reduced 2-component system deserves a special consideration. In particular, we show by establishing a Lax pair that the system is completely integrable.
The properties of solutions such as loop solitons and breathers are investigated in detail, confirming their solitonic behavior.
A variant of the 2-component system is also discussed with its multisoliton solutions.

\newpage
\leftline{\bf  I. INTRODUCTION} \par
The short pulse (SP) equation was derived as a model nonlinear equation 
describing the propagation of ultra-short  pulses in isotropic optical fibers.$^{1}$
We write it in an appropriate dimensionless form as
$$u_{xt}=u+{1\over 6 }(u^3)_{xx}, \eqno(1.1)$$
where $u=u(x,t)$ represents the magnitude of the electric field and subscripts $x$ and $t$
appended to $u$ denote  partial differentiations.
The SP equation has appeared for the first time
in an attempt to construct integrable differential equations associated with pseudospherical surfaces.$^{2}$
The integrability, soliton solutions and other features  of the SP equation common to the completely integrable
partial differential equations (PDEs) have been studied from various points 
of view.$^{2-10}$ See also Ref. 11 for a recent review article on the SP equation
which is mainly concerned with soliton and periodic solutions and their properties.
It also provides a novel method for constructing multiperiodic solutions by means of the bilinear
transformation method. \par
There exist a few generalizations of the SP equation to the two-component systems
that take into account the effects of polarization and nonisotropy. One is due to Pietrzyk {\it et al}.
They proposed the following three integrable vector (or two-component) SP equations:$^{12}$
$$u_{xt}=u+{1\over 6}(u^3+3uv^2)_{xx},\quad v_{xt}=v+{1\over 6}(v^3+3u^2v)_{xx}, \eqno(1.2)$$
$$u_{xt}=u+{1\over 6}(u^3-3uv^2)_{xx},\quad v_{xt}=v-{1\over 6}(v^3-3u^2v)_{xx}, \eqno(1.3)$$
$$u_{xt}=u+{1\over 6}(u^3)_{xx},\quad v_{xt}=v+{1\over 2}(u^2v)_{xx}. \eqno(1.4)$$
Another one is given by Sakovich:$^{13}$
$$u_{xt}=u+{1\over 6}(u^3+uv^2)_{xx},\quad v_{xt}=v+{1\over 6}(v^3+u^2v)_{xx}, \eqno(1.5)$$
$$u_{xt}=u+{1\over 6}(u^3)_{xx},\quad v_{xt}=v+{1\over 6}(u^2v)_{xx}. \eqno(1.6)$$
As pointed out by Sakovich,$^{13}$ the two systems (1.2) and (1.3) can be reduced to the SP equation (1.1)
by  appropriate dependent variable transformations. Indeed, introducing the new variables $p$ and $q$ by $p=u+v, q=u-v$,
the system of equations (1.2) can be decoupled and both $p$ and $q$ satisfy the SP equation (1.1) while for (1.3), the
transformation $p=u+{\rm i}v$ and $q=u-{\rm i}v$ leads to the two  decoupled SP equations as well. 
On the other hand, the integrability of the systems (1.5) and (1.6)
was investigated by means of the Painlev\'e analysis. Sakovich showed that the above two systems pass the Painlev\'e
test, providing a strong indication of their integrability. Nevertheless, their Lax representations, conservations laws and soliton solutions
have not been obtained as yet for the systems. \par
The purpose of this paper is to propose a novel multi-component analog of the SP equation and construct its multisoliton solutions.
The system of equations presented here is composed of the following coupled nonlinear PDEs for the $n$ variables $u_i (i=1, 2, ..., n)$:
$$u_{i,xt}=u_i+{1\over 2}(Fu_{i,x})_x, \quad (i=1, 2, ..., n), \eqno(1.7a)$$
with
$$ \quad F=\sum_{1\leq j<k\leq n}c_{jk}u_ju_k. \eqno(1.7b)$$
Here, $c_{jk}$ are arbitrary constants with the symmetry $c_{jk}=c_{kj}  (j, k= 1,2, ..., n).$
For the special case of $n=2$ with $c_{12}=1$, this system becomes
$$u_{xt}=u+{1\over 2}(uvu_x)_x,\quad v_{xt}=v+{1\over 2}(uvv_x)_x, \eqno(1.8)$$
where $u=u_1$ and $v=u_2$.
Obviously, if we put $u=v$, then (1.8)  reduces to the SP equation (1.1).
A simple transformation recasts (1.8) to the system of equations 
$$u_{xt}=u+{1\over 2}[(u^2+v^2)u_x]_x,\quad v_{xt}=v+{1\over 2}[(u^2+v^2)v_x]_x. \eqno(1.9)$$
If $v=0$, then this system reduces to the SP equation (1.1).
  The present paper is organized as follows. In Sec. II, we summarize an exact method of solution for the
SP equation which will be suitable for application to the multi-component system.
 In Sec. III,  We show by applying  the standard procedure of the bilinear method that the system of equations (1.7) can be
transformed to a coupled system of bilinear equations and obtain the multisoliton soliton solution in the parametric form. 
Notably, the tau-functions constituting the solution are expressed in terms of pfaffians unlike the determinantal solutions of
the SP equation.$^{9}$
The proof of the multisoliton solution is, however,  performed with use of an elementary theory of determinants without recourse to the pfaffian theory.
 In Section IV, we consider  the system (1.8).
 In particular, we demonstrate that it is a completely integrable system by establishing a Lax pair. 
 The multisoliton solution to the system is reduced from that of the $n$-component system.  
 The properties of the 1- and 2-soliton solutions will be investigated in detail. 
 Subsequently, we  briefly discuss the system (1.9).
  In Sec V, we conclude this study with a short summary and discuss some open problems associated with the multi-component SP equations. 
 \par
\bigskip
\noindent {\bf II. SUMMARY OF THE EXACT METHOD OF SOLUTION} \par
Here, we give a short summary of the exact method of solution for the SP equation. 
Although we have employed some nonlinear transformations to reduce the SP to the integrable sine-Gordon (sG)
equations,$^{4,9,10}$ we provide a different approach which is more suitable for solving
 the system of equations (1.7). \par
 \bigskip
 \noindent{\bf A. Hodograph transformation}\par
 We first introduce the hodograph transformation $(x ,t) \rightarrow (y, \tau)$ by
$$dy=rdx+{1\over 2}u^2rdt, \quad d\tau=dt, \eqno(2.1a)$$
where $r(>0)$ is a function of $u$ to be determined later.
Using (2.1a), the $x$ and $t$ derivatives are rewritten as
$${\partial\over\partial x}=r{\partial\over\partial y},
\quad {\partial\over\partial t}={\partial\over\partial \tau}+{1\over 2}u^2r{\partial\over\partial y}. \eqno(2.1b)$$
It follows from (2.1b) that $x=x(y, \tau)$ satisfies the system of linear PDEs
$$x_y={1\over r},\quad x_\tau=-{u^2\over 2}. \eqno(2.2)$$
Equation (1.1) is then transformed into the form
$$u_{y\tau}=x_yu. \eqno(2.3)$$
 The form of $r$ can be determined by the solvability condition of the system (2.2), i.e.,
$x_{y\tau}=x_{\tau y}$. Indeed, this immediately gives  $r_\tau=uu_yr^2$.
On the other hand, it follows from (2.2) and (2.3) that $u=ru_{y\tau}$. Eliminating the variable $u$ from both relations, one has $r_\tau=u_yu_{y\tau}r^3$.
  If we impose the boundary conditions $u(\pm\infty, \tau)=0, r(\pm\infty, \tau)=1$, then we obtain $r^2=(1-u_y^2)^{-1}$ after
  integrating this relation with respect to $\tau$. 
Since $u_y=u_x/r$ by (2.1b), we can rewrite this expression into the form
$$r^2=1+u_x^2. \eqno(2.4)$$
The above relation has been used to transform the SP equation into the  form of conservation law $r_t=(u^2r/2)_x$. 
If one introduces a new variable $\phi$ by $u_y=\sin\,\phi$, then $\phi$ satisfies the sG equation
$\phi_{y\tau}=\sin\,\phi$. This equation was the starting point in constructing multisoliton solutions of the SP equation.$^{9}$
Below, we develop an alternative method using (2.3) which will be relevant to application to the multi-component system.
 \par
\bigskip
\noindent{\bf B. Parametric representation of soliton solutions} \par
The soliton solutions of Eq. (2.3) are constructed by a direct method using the bilinear formalism.
To this end, we first introduce the following dependent variable transformations for $u$ and $x$
$$u={g\over f}, \eqno(2.5)$$
$$x=y+{h\over f}, \eqno(2.6)$$
where $f$, $g$ and $h$ are tau-functions. Note that we may add an arbitrary constant on the right-hand side of (2.6), if necessary.
The second equation of (2.2) is then transformed to the the bilinear equation
$$2D_\tau h\cdot f+g^2=0, \eqno(2.7)$$
where the bilinear operators $D_\tau$ and $D_y$ are defined by
$$D_\tau^mD_y^nf\cdot g=\left({\partial\over\partial \tau}-{\partial\over\partial \tau^\prime}\right)^m
\left({\partial\over\partial y}-{\partial\over\partial y^\prime}\right)^n
f(\tau, y)g(\tau^\prime, y^\prime)\Big|_{\tau^\prime=\tau,\, y^\prime=y},\quad (m, n = 0, 1, 2, ...).  \eqno(2.8)$$
On the other hand,  Eq. (2.3) becomes
$${gf_y\over f^3}(2f_\tau+h)-{1\over f^2}(f_\tau g_y+f_yg_\tau+f_{y\tau}g+gh_y)+{1\over f}(g_{y\tau}-g)=0. \eqno(2.9)$$
We can decouple Eq. (2.9) to a set of equations
$$2f_\tau+h=0, \eqno(2.10)$$
$$f_\tau g_y+f_yg_\tau+f_{y\tau}g+gh_y-f(g_{y\tau}-g)=0. \eqno(2.11)$$
Substituting $h$ from (2.10) into Eqs. (2.11) and (2.7), we obtain the following system of bilinear equations for $f$ and $g$:
$$D_yD_\tau f\cdot g=fg, \eqno(2.12)$$
$$D_\tau^2f\cdot f={1\over 2}\,g^2. \eqno(2.13)$$
It then follows from (2.6) and (2.10) that
$$x=y-2\,{f_\tau\over f}. \eqno(2.14)$$
Thus, the soliton solutions of the SP equation are given by the parametric representation (2.5) and (2.14) in terms
of the tau-functions $f$ and $g$. In the simplest case of the 1-soliton solution, the solutions to Eqs. (2.12) and (2.13) are easily found to be as
$$f=1+{\rm e}^{2\xi},\quad g={4\over p}\,{\rm e}^\xi,\quad \xi=py+{1\over p}\tau +\xi_0, \eqno(2.15)$$
where $p$ and $\xi_0$ are constants related to the amplitude and phase of the soliton, respectively. The corresponding parametric 
representation of the 1-soliton solution is  derived from (2.5), (2.14) and (2.15). It reads  
$$u={2\over p}\,{\rm sech\,\xi},\quad x=y-{2\over p}\,\tanh\,\xi +x_0, \eqno(2.16)$$
where $x_0=-p/2$. For real $p$ and $\xi_0$, the solution takes the form of a loop soliton.\par
\bigskip
\noindent{\bf C. Remark}\par
We have already shown that the SP equation can be transformed into the sG equation and obtained the 
parametric representation of the $N$-soliton solution. Actually, it reads$^{9}$
$$u= 2{\rm i}\left(\ln{\tilde f^\prime\over \tilde f}\right)_\tau,\quad x=y-2\,(\ln\,{\tilde f}^\prime\tilde f)_\tau, \eqno(2.17)$$
where $\tilde f$ and ${\tilde f}^\prime$ are tau-functions for the sG equation $\phi_{y\tau}=\sin\,\phi,\ \phi=2{\rm i}\,\ln({\tilde f}^\prime/\tilde f)$
and they satisfy the bilinear equations
$$D_yD_\tau {\tilde f}\cdot {\tilde f}={1\over 2}(\tilde f^2-{\tilde f}^{\prime 2}),
\quad D_yD_\tau {\tilde f}^\prime\cdot {\tilde f}^\prime={1\over 2}({\tilde f}^{\prime 2}-\tilde f^2). \eqno(2.18)$$
The explict forms of the tau-functions are given by
$$\tilde f=\sum_{\mu=0,1}{\rm exp}\left[\sum_{j=1}^N\mu_j\left(\xi_j+{\pi\over 2}{\rm i}\right)
+\sum_{1\le j<k\le N}\mu_j\mu_k\gamma_{jk}\right], \eqno(2.19a)$$
  $$\tilde f^\prime=\sum_{\mu=0,1}{\rm exp}\left[\sum_{j=1}^N\mu_j\left(\xi_j-{\pi\over 2}{\rm i}\right)
+\sum_{1\le j<k\le N}\mu_j\mu_k\gamma_{jk}\right], \eqno(2.19b)$$
where
$$\xi_j=p_jy+{1\over p_j}\tau+\xi_{j0}, \quad (j=1, 2, ..., N),\eqno(2.20a)$$
$$e^{\gamma_{jk}}=\left({p_j-p_k\over p_j+p_k}\right)^2, \quad (j, k=1, 2, ..., N; j\not=k).
\eqno(2.20b)$$
Here, $p_j$ and $\xi_{j0}$ are arbitrary complex-valued  parameters satisfying the conditions $p_j\not=\pm p_k$
for $j\not= k$ and $N$ is an arbitrary positive integer. The notation $\sum_{\mu=0,1}$
implies the summation over all possible combination of $\mu_1=0, 1, \mu_2=0, 1, ..., 
\mu_N=0, 1$.
Thus, we have two different expressions for the parametric soliton solutions of the SP equation, i.e., one
is (2.5) with (2.14) and the other is (2.17).
We can show that the tau-functions $f$ and $g$ are related to the
tau-functions $\tilde f$ and $\tilde f^\prime$ by the relations
$$f=\tilde f^\prime\tilde f, \quad g= 2{\rm i}\,D_\tau\tilde f^\prime\cdot \tilde f, \eqno(2.21)$$
which will be inferred by comparing (2.5) and (2.14) with (2.17). \par
\bigskip
\noindent{\bf III. MULTI-COMPONENT SYSTEM}\par
Let us now consider the multi-component system (1.7). The procedure for obtaining the parametric representation of 
soliton solutions parallels that developed in Sec. II for the SP equation.  Hence, we omit the detail of the derivation and write down the final results.
Specifically, we give the parametric representation of soliton solutions and associated system of bilinear equations corresponding to Eqs. (2.12) and (2.13).
Then, we present the explicit form of the multisoliton solution of the bilinear equations. Last, the proof of the multisoliton solution is performed by using
an elementary theory of determinants. \par
\noindent{\bf A. Parametric representation of soliton solutions}\par
If we use the hodograph transformation (2.1a) with $F$ given by (1.7b) in place of $u^2$
$$dy=rdx+{1\over 2}Frdt, \quad d\tau=dt, \eqno(3.1)$$
we then obtain the equations corresponding to Eqs. (2.2) and (2.3) which are given respectively by
$$x_y={1\over r},\quad x_\tau=-{F\over 2}. \eqno(3.2)$$
$$u_{i,y\tau}=x_yu_i \quad (i= 1, 2, ..., n). \eqno(3.3)$$
The solvability condition for Eqs. (3.2) gives the explicit form of $r^2$ in terms of $u_{i,y}\,(i= 1, 2, ..., n)$ as
$$r^2={1\over 1-\sum_{1\leq j<k\leq n}c_{jk}u_{j,y}u_{k,y}}. \eqno(3.4a)$$
If we use the relation $u_{j,y}=u_{j,x}/r$, then we can rewrite (3.4a) in terms of the original variable $u_{i,x}\,(i= 1, 2, ..., n)$
$$r^2=1+\sum_{1\leq j<k\leq n}c_{jk}u_{j,x}u_{k,x}. \eqno(3.4b)$$
The parametric representation of the soliton solutions takes the form
$$u_i={g_i\over f},\quad (i= 1, 2, ..., n), \quad x=y-2\,{f_\tau\over f}, \eqno(3.5)$$
where the tau-functions $f$ and $g_i (i= 1,2, ..., n)$ satisfy the system of bilinear equations
$$D_yD_\tau f\cdot g_i=fg_i, \quad (i= 1, 2, ..., n), \eqno(3.6)$$
$$D_\tau^2f\cdot f={1\over 2\,}\sum_{1\leq j<k\leq n}c_{jk}g_jg_k. \eqno(3.7)$$    
It follows from (3.2)-(3.4a) that $u_i=u_i(y, \tau)$ obey a closed system of PDEs
$${u_{i,y\tau}\over \sqrt{1-\sum_{1\leq j<k\leq n}c_{jk}u_{j,y}u_{k,y}}}=u_i,\quad (i= 1, 2, ..., n). \eqno(3.8)$$
Furthermore, if we introduce the new variables $v_i$ by $v_i=u_{i,y}\,(i= 1, 2, ..., n)$, then the above system can be recast to
$${\partial \over \partial y}\left[{v_{i,\tau}\over \sqrt{1-\sum_{1\leq j<k\leq n}c_{jk}v_jv_k}}\right]=v_i,\quad (i= 1, 2, ..., n). \eqno(3.9)$$
\bigskip
\noindent {\bf B. Multisoliton solution of bilinear equations} \par
We first introduce vectors and matrices.    
Subsequently, we present the explicit multisoliton solution of the bilinear equations (3.6) and (3.7).
\par
\noindent{\bf 1. Definition} \par
 Let ${\bf a, b, c}$ and ${\bf 0}$ be 
row vectors having $M$ components 
$${\bf a}=(a_1, a_2, ..., a_M), \quad {\bf b}=(b_1, b_2, ..., b_M), \quad {\bf c}=(c_1, c_2, ..., c_M),$$
$$ \quad {\bf d}=(d_1, d_2, ..., d_M), \quad {\bf 0}=(0, 0, ..., 0), \eqno(3.10a)$$
and ${\bf e}_i\ (i=1, 2, ..., n)$ be $M$-component row vectors defined below
$${\bf e}_1=(\underbrace{1, 1, ..., 1}_{M_1},\underbrace{0, 0, ..., 0}_{M-M_1}), ..., 
{\bf e}_i=(\underbrace{0, 0, ..., 0}_{ M_1+\cdots +M_{i-1}}, \underbrace{1, 1, ..., 1}_{M_i}, \underbrace{0, 0, ..., 0}_{M-(M_1+\cdots +M_i)}),$$
 $$..., {\bf e}_n=(\underbrace{0, 0, ..., 0}_{M_1+\cdots+M_{n-1}}, \underbrace{1, 1, ..., 1}_{M_n}), \eqno(3.10b)$$
where $M$ and $M_i (i=1, 2, ..., n)$ are positive integers satisfying the condition $\sum_{i=1}^nM_i=M$.
\par
 The following types of matrices appear in the process of proving the multisoliton solution:
$$D=(d_{ij})_{1\leq i,j\leq 2M}=\begin{pmatrix} A_M & I_M \\ -I_M & B_M \end{pmatrix}, 
\quad D({\bf a}; {\bf b})=\begin{pmatrix} A_M & I_M &{\bf b}^T\\ -I_M & B_M & {\bf 0}^T \\ {\bf a}& {\bf 0} &0 \end{pmatrix}, \eqno(3.11a)$$
$$D({\bf a}, {\bf b}; {\bf c}, {\bf d})=\begin{pmatrix} A_M & I_M & {\bf c}^T & {\bf d}^T \\
                                                       -I_M & B_M & {\bf 0}^T & {\bf 0}^T \\
                                                        {\bf a} & {\bf 0} &  0 & 0 \\
                                                        {\bf b} & {\bf 0} &  0 & 0 \end{pmatrix}, \quad
  D({\bf a}, {\bf e}_i; {\bf b}, {\bf e}_j)=\begin{pmatrix} A_M & I_M & {\bf b}^T & {\bf 0}^T \\
                                                       -I_M & B_M & {\bf 0}^T & {\bf e}_j^T \\
                                                        {\bf a} & {\bf 0} &  0 & 0 \\
                                                        {\bf 0} & {\bf e}_i &  0 & 0 \end{pmatrix}, \eqno(3.11b)$$ 
$$D({\bf a}, {\bf b}, {\bf e}_i; {\bf c}, {\bf d}, {\bf e}_j)=\begin{pmatrix} A_M & I_M & {\bf c}^T & {\bf d}^T & {\bf 0}^T  \\
                                                                             -I_M & B_M & {\bf 0}^T & {\bf 0}^T & {\bf e}_j^T \\
                                                                            {\bf a} & {\bf 0} &  0 & 0 & 0 \\
                                                                            {\bf b} & {\bf 0} &  0 & 0 & 0 \\
                                                                            {\bf 0} & {\bf e}_i &  0 & 0& 0  \end{pmatrix}, \eqno(3.11c)$$
where $A_M=(a_{ij})_{1\leq i,j\leq M}$ and $B_M=(b_{ij})_{1\leq i,j\leq M}$ are $M\times M$ skew-symmetic matrices, $I_M$ is the $M\times M$ unit matrix and
the symbol $T$ denotes the transpose.\par
The element $a_{ij}$ of the matrix $A_M$ is given by
$$a_{ij}=-{p_i-p_j\over p_i+p_j}\,{\rm e}^{\xi_i+\xi_j}=-{p_i-p_j\over p_i+p_j}z_iz_j,\ (i, j= 1, 2, ..., M), \eqno(3.12)$$
where $\xi_i$ is defined by (2.20a) and $z_i={\rm e}^{\xi_i}$.
To specify the matrix $B_M$, let $S_i\, (i=1, 2, ..., n)$ be $n$ disjoint sets consisting of positive integers
$$S_1=\{1,..., M_1\}, ..., S_i=\{M_1+M_2+\cdots +M_{i-1}+1, ..., M_1+\cdots +M_i\},$$
$$ ..., S_n=\{M_1+M_2+\cdots +M_{n-1}+1, ..., M_1+\cdots +M_n\}. \eqno(3.13)$$
Then
$$b_{\mu\nu}={1\over 4\,}c_{ij}\,{(p_\mu p_\nu)^2\over p_\mu^2-p_\nu^2},\ \mu\in S_i, \nu\in S_j\ (\mu, \nu=1, 2, ..., M\ (\mu\not=\nu);
i, j=1, 2, ..., n\ (i\not=j)), \eqno(3.14)$$
 $b_{\mu\nu}=0$ if $\mu$ and $\nu$ belong to the same set and $b_{\mu\mu}=0$ for all $\mu$. Thus, $B_M$ has the structure
$$B_M=\begin{pmatrix} O_{M_1\times M_1} & B_{M_1\times M_2} & ... & B_{M_1\times M_n} \\
-B^T_{M_1\times M_2} & O_{M_2\times M_2} & ... & B_{M_2\times M_n} \\
\vdots & \vdots & \ddots & \vdots \\
-B^T_{M_1\times M_n} & -B^T_{M_2\times M_n} & ... & O_{M_n\times M_n}\end{pmatrix}, \eqno(3.15a)$$
$$B_{M_i\times M_j}=(b_{\mu\nu})_{\mu\in S_i, \nu\in S_j}\ (1\leq i<j\leq n),$$
$$\quad O_{M_i\times M_i}: M_i\times M_i\ {\rm null\ matrix}\ (i=1, 2, ..., n). \eqno(3.15b)$$
\par
\noindent {\bf 2. Multisoliton solution}\par
Now, we state our main result. \par
{\bf Theorem 3.1:} {\it The multisoliton solution of the system of bilinear equations (3.6) and (3.7) is given by the following form:
$$f=\sqrt{F}, \quad F=|D|, \eqno(3.16a)$$
$$g_i=\sqrt{G_i},\quad G_i=|D(-{\bf z},-{\bf e}_i; {\bf z},{\bf e}_i)|, \ (i=1, 2, ..., n), \eqno(3.16b)$$
where ${\bf z}$ is the $M$-component vector ${\bf z}=({\rm e}^{\xi_1}, {\rm e}^{\xi_2}, ..., {\rm e}^{\xi_M})$.
The parametric solution $u_i$  (3.5) constructed from these tau-functions contains $M_i$ solitons for each $i$.} \par
Note that $f$ and $g_i$ are pfaffians since each one of them is represented by the square root of the skew-symmetric determinant of even oder.
This fact is in striking contrast to the tau-functions of the $N$-soliton solution for the SP equation which can be
represented by determinants.
\par
\bigskip
\noindent{\bf C. PROOF OF MULTISOLITON SOLUTION}\par
\noindent{\bf 1. Basic formulas for determinants}\par
Let $A=(a_{ij})_{1\leq i,j\leq M}$ be an $M\times M$ matrix and $A_{ij}$ be the cofactor of the element $a_{ij}$. Then,  the following three formulas for determinants
are employed frequently in our analysis:$^{14}$
$${\partial\over\partial x}|A|=\sum_{i,j=1}^M{\partial a_{ij}\over\partial x}A_{ij}, \eqno(3.17)$$
$$\begin{vmatrix} A & {\bf a}^T\\ {\bf b} & z\end{vmatrix}=|A|z-\sum_{i,j=1}^MA_{ij}a_ib_j,  \eqno(3.18)$$
$$|A({\bf a}, {\bf b}; {\bf c}, {\bf d})||A|= |A({\bf a}; {\bf c})||A({\bf b}; {\bf d})|-|A({\bf a}; {\bf d})||A({\bf b}; {\bf c})|. \eqno(3.19)$$
The formula (3.17) is the differential rule of the determinant and (3.18) is the expansion formula for a bordered determinant
with respect to the last row and column.
The formula (3.19) is Jacobi's identity and it will play a central role in the proof of the multisoliton solution. \par
\noindent{\bf 2. Differential formulas } \par
We give various differential formulas for the determinants $F$ and $G_i$ introduced in (3.16) which are 
necessary for the proof of the solution. The following formulas are derived easily with use of (3.17) and (3.18)
as well as the relation $|D(-{\bf z}; {\bf z})|=0$ which follows from the fact that the skew-symmetric determinant of odd order
is identically zero. Hence, we quote only the results:
$$F_y=-2|D(-{\bf z}; {\bf z}_y)|, \eqno(3.20a)$$
$$F_\tau=-2|D(-{\bf z}_\tau; {\bf z})|, \eqno(3.20b)$$
$$F_{y\tau}=-2|D(-{\bf z}_\tau; {\bf z}_y)|-2|D(-{\bf z},-{\bf z}_\tau; {\bf z}, {\bf z}_y)|, \eqno(3.20c)$$
$$F_{\tau\tau}=-2|D(-{\bf z}_{\tau\tau}; {\bf z})|-2|D(-{\bf z},-{\bf z}_\tau; {\bf z}, {\bf z}_\tau)|, \eqno(3.20d)$$
$$G_{i,y}=2|D(-{\bf z},-{\bf e}_i; {\bf z}_y,{\bf e}_i)|, \eqno(3.21a)$$
$$G_{i,\tau}=2|D(-{\bf z}_\tau,-{\bf e}_i; {\bf z},{\bf e}_i)|, \eqno(3.21b)$$
$$G_{i,y\tau}=2|D(-{\bf z},-{\bf e}_i; {\bf z},{\bf e}_i)|+2|D(-{\bf z}_\tau,-{\bf e}_i; {\bf z}_y,{\bf e}_i)|
+2|D(-{\bf z},-{\bf z}_\tau,-{\bf e}_i; {\bf z},{\bf z}_y,{\bf e}_i)|, \eqno(3.21c)$$
where the $M$-component vectors ${\bf z}_y, {\bf z}_\tau$ and ${\bf z}_{\tau\tau} $ are given respectively  by
$${\bf z}_y=(p_1{\rm e}^{\xi_1}, p_2{\rm e}^{\xi_2}, ..., p_M{\rm e}^{\xi_M}), 
\quad {\bf z}_\tau=\left({{\rm e}^{\xi_1}\over p_1}, {{\rm e}^{\xi_2}\over p_2}, ..., {{\rm e}^{\xi_M}\over p_M}\right),
\quad {\bf z}_{\tau\tau}=\left({{\rm e}^{\xi_1}\over p_1^2}, {{\rm e}^{\xi_2}\over p_2^2}, ..., {{\rm e}^{\xi_M}\over p_M^2}\right).\eqno(3.22)$$
\noindent{\bf 3. Proof of Eq. (3.6) } \par
First, we show that the tau-functions (3.16) for the multisoliton solution satisfy the bilinear equation (3.6).  To this end, 
we substitute $f$ and $g_i$ from (3.16) into Eq. (3.6) to obtain
$${G_i\over 2F}\left(FF_{y\tau}-{1\over 2}F_yF_\tau\right)+{F\over 2G_i}\left(G_iG_{i,y\tau}-{1\over2}G_{i,y}G_{i,\tau}\right)
-{1\over 4}(F_yG_{i,\tau}+F_\tau G_{i,y})=FG_i. \eqno(3.23)$$
We compute three terms on the left-hand side of (3.23) separately. Using (3.20a)-(3.20c) and the relation
$$|D(-{\bf z},-{\bf z}_\tau; {\bf z}, {\bf z}_y)||D|=-|D(-{\bf z}; {\bf z}_y)||D(-{\bf z}_\tau; {\bf z})|, \eqno(3.24)$$
which follows from Jacobi's identity and the identity $|D(-{\bf z}; {\bf z})|=0$, the first term on the left-hand side of (3.23) reduces to
$$ {G_i\over 2F}\left(FF_{y\tau}-{1\over 2}F_yF_\tau\right)=-|D(-{\bf z}_\tau; {\bf z}_y)|G_i. \eqno(3.25)$$
\par
Next, it follows from (3.21a)-(3.21c) that
$$G_iG_{i,y\tau}-{1\over2}G_{i,y}G_{i,\tau}=2|D(-{\bf z},-{\bf e}_i; {\bf z},{\bf e}_i)|\Bigl\{
|D(-{\bf z},-{\bf e}_i; {\bf z},{\bf e}_i)|+|D(-{\bf z}_\tau,-{\bf e}_i; {\bf z}_y,{\bf e}_i)|$$
$$+ |D(-{\bf z},-{\bf z}_\tau,-{\bf e}_i; {\bf z},{\bf z}_y,{\bf e}_i)|\Bigr\}
-2|D(-{\bf z},-{\bf e}_i; {\bf z}_y,{\bf e}_i)||D(-{\bf z}_\tau,-{\bf e}_i; {\bf z},{\bf e}_i)|. \eqno(3.26)$$
Referring again to Jacobi's identity and the identity $|D(-{\bf e}_i; {\bf e}_i)|=0$, one has
$$|D(-{\bf z},-{\bf z}_\tau,-{\bf e}_i; {\bf z},{\bf z}_y,{\bf e}_i)||D(-{\bf e}_i; {\bf e}_i)|$$
$$=|D(-{\bf z},-{\bf e}_i; {\bf z},{\bf e}_i)||D(-{\bf z}_\tau,-{\bf e}_i; {\bf z}_y,{\bf e}_i)|
-|D(-{\bf z}_\tau,-{\bf e}_i; {\bf z},{\bf e}_i)||D(-{\bf z},-{\bf e}_i; {\bf z}_y,{\bf e}_i)|=0, \eqno(3.27)$$
which, introduced into (3.26), simplifies the second term on the left-hand side of (3.23)
$${F\over 2G_i}\left(G_iG_{i,y\tau}-{1\over2}G_{i,y}G_{i,\tau}\right)=F\Bigl\{|D(-{\bf z},-{\bf z}_\tau,-{\bf e}_i; {\bf z},{\bf z}_y,{\bf e}_i)|+G_i\Bigr\}. \eqno(3.28)$$
\par
Last, the formulas (3.20a), (3.20b), (3.21a) and (3.21b) give simply the third term on the left-hand side of (3.23)
$$-{1\over 4}(F_yG_{i,\tau}+F_\tau G_{i,y})=|D(-{\bf z}; {\bf z}_y)||D(-{\bf z}_\tau,-{\bf e}_i; {\bf z},{\bf e}_i)|
+|D(-{\bf z}_\tau; {\bf z})||D(-{\bf z},-{\bf e}_i; {\bf z}_y,{\bf e}_i)|. \eqno(3.29)$$
Substituting (3.25), (3.28) and (3.29) into (3.23), the equation to be proved becomes
$$|D||D(-{\bf z},-{\bf z}_\tau,-{\bf e}_i; {\bf z},{\bf z}_y,{\bf e}_i)|
-|D(-{\bf z}_\tau; {\bf z}_y)||D(-{\bf z},-{\bf e}_i; {\bf z},{\bf e}_i)|$$
$$ +|D(-{\bf z}; {\bf z}_y)||D(-{\bf z}_\tau,-{\bf e}_i; {\bf z},{\bf e}_i)|
+|D(-{\bf z}_\tau; {\bf z})||D(-{\bf z},-{\bf e}_i; {\bf z}_y,{\bf e}_i)|=0. \eqno(3.30)$$
The following formula can be verified  by applying Jacobi's identity twice to the right-hand side of (3.31):
$$ \begin{vmatrix} |D({\bf a}; {\bf a}^\prime)| &|D({\bf a}; {\bf b}^\prime)|&|D({\bf a}; {\bf c}^\prime)| \\
                   |D({\bf b}; {\bf a}^\prime)| &|D({\bf b}; {\bf b}^\prime)|&|D({\bf b}; {\bf c}^\prime)| \\
                   |D({\bf c}; {\bf a}^\prime)| &|D({\bf c}; {\bf b}^\prime)|&|D({\bf c}; {\bf c}^\prime)| \end{vmatrix}
                   =|D|^2|D({\bf a},{\bf b},{\bf c}; {\bf a}^\prime,{\bf b}^\prime,{\bf c}^\prime)|. \eqno(3.31)$$
Assume that $|D|\not=0$. Then, multiplying (3.30) by $|D|$ and using Jacobi's identity as well as the identities
$|D(-{\bf e}_i; {\bf e}_i)|=|D(-{\bf z}; {\bf z})|=0$, the resulting relation  reduces to (3.31) with the
identification ${\bf a}=-{\bf z}, {\bf b}=-{\bf z}_\tau, {\bf c}=-{\bf e}_i, {\bf a}^\prime={\bf z}, {\bf b}^\prime={\bf z}_y, {\bf c}^\prime={\bf e}_i$.
This completes the proof of Eq. (3.6). \par
\noindent{\bf 4. Proof of Eq. (3.7) } \par
We proceed to the proof of Eq. (3.7).  By using $f$ and $g_i$ from (3.16) and noting the symmetry $c_{ij}=c_{ji}$, we transform it to the form
$$FF_{\tau\tau}-F_\tau^2={1\over 4}\sum_{\substack{j,k=1\\ (j\not=k)}}^nc_{jk} G_jG_k. \eqno(3.32)$$
If we substitute (3.16b), (3.20b) and (3.20d) into (3.32) and use the following relation with $j=k$ 
$$|D(-{\bf e}_j; {\bf z})||D(-{\bf e}_k; {\bf z})|=|D||D(-{\bf z},-{\bf e}_j; {\bf z},{\bf e}_k)|,\ (j, k=1, 2, ..., n), \eqno(3.33)$$
which comes from Jacobi's identity, we recast (3.32) in the form
$$2|D|\Bigl\{|D(-{\bf z}; {\bf z}_{\tau\tau})|-|D(-{\bf z},-{\bf z}_\tau; {\bf z}, {\bf z}_\tau)|\Bigr\}
={1\over 4}\sum_{\substack{j,k=1\\ (j\not=k)}}^nc_{jk}|D(-{\bf e}_j; {\bf z})||D(-{\bf e}_k; {\bf z})|. \eqno(3.34)$$
Last,  replacing the right-hand side of (3.34) by the right-hand side of (3.33) and dividing the resultant equation by $2|D|$, the equation to be proved reduces to
the following {\it linear} relation among determinants:
$$|D(-{\bf z}; {\bf z}_{\tau\tau})|-|D(-{\bf z},-{\bf z}_\tau; {\bf z}, {\bf z}_\tau)|
={1\over 8}\sum_{\substack{j,k=1\\ (j\not=k)}}^nc_{jk}|D(-{\bf z},-{\bf e}_j; {\bf z},{\bf e}_k)|. \eqno(3.35)$$
\par
We now start the proof of (3.35).
Define the $(2M+1)\times(2M+1)$ skew-symmetric matrix $D^\prime=(d_{ij}^\prime)_{1\leq i,j\leq 2M+1}$ by
$$D^\prime=D(-{\bf z}; {\bf z})=\begin{pmatrix} A_M & I_M &{\bf z}^T\\ -I_M & B_M & {\bf 0}^T \\ -{\bf z}& {\bf 0} &0 \end{pmatrix}. \eqno(3.36)$$
Let $D_{ij}$ and $D_{ij}^\prime$ be the cofactors of the elements $d_{ij}$ and $d_{ij}^\prime$, respectively and
$D_{ij,kl}$ and $D_{ij,kl}^\prime$ be second cofactors. Expanding the cofactor $D_{M+j,M+i}^\prime$ with respect to $i$th row, we obtain
$$D_{M+j,M+i}^\prime =\sum_{k=1}^MD_{i\,M+j,k\,M+i}^\prime a_{ik}+\sum_{k=1}^MD_{i\,M+j,k\,M+i}\,z_iz_k,\ (i, j=1, 2, ..., M). \eqno(3.37)$$
Similarly, referring to the structure of the matrix $B_M$ defined by (3.15), the expansions of $D_{ij}$ and $D_{ij}^\prime$ with respect to the
$(M+i)$th column read
$$D_{ij}=\sum_{k=1}^MD_{i\,M+k,j\,M+i}b_{ki},\ (i, j=1, 2, ..., M), \eqno(3.38)$$
$$D_{ij}^\prime=\sum_{k=1}^MD_{i\,M+k,j\,M+i}^\prime b_{ki},\ (i, j=1, 2, ..., M). \eqno(3.39)$$
The proof of (3.35) can be performed on the basis of the formulas (3.37)-(3.39).
First, we multiply (3.37) by $b_{ji}/p_i^2$ and sum up with respect to $i$ and $j$ to obtain
$$\sum_{i,j=1}^MD_{M+j,M+i}^\prime {b_{ji}\over p_i^2}=\sum_{i,j=1}^M\sum_{k=1}^MD_{i\,M+j,k\,M+i}^\prime a_{ik}{b_{ji}\over p_i^2}$$
$$+\sum_{i,j=1}^M\sum_{k=1}^MD_{i\,M+j,k\,M+i}{b_{ji}\over p_i^2}z_iz_k,\ (i, j=1, 2, ..., M). \eqno(3.40)$$
Note that for any function $f_{ij}$
$$\sum_{i,j=1}^Mf_{ij}=\sum_{i,j=1}^n\sum_{\mu\in S_i}\sum_{\nu\in S_j}f_{\mu\nu}, \eqno(3.41)$$
where the notation $\sum_{\mu\in S_i}$ implies that the dummy index $\mu$ runs over the set $S_i$.
Applying this rule to the left-hand side of (3.40)
$$L\equiv \sum_{i,j=1}^MD_{M+j,M+i}^\prime {b_{ji}\over p_i^2} = \sum_{i,j=1}^n\sum_{\mu\in S_i}\sum_{\nu\in S_j}D_{M+\nu,M+\mu}^\prime {b_{\nu\mu}\over p_\mu^2}. \eqno(3.42)$$
We modify $L$ by taking into account the relations $b_{\nu\mu}=- b_{\mu\nu}$ and $D_{M+\nu,M+\mu}^\prime=D_{M+\mu,M+\nu}^\prime$ which follow from the skew-symmetry of the
matrics $D$ and $D
^\prime$.  This leads to
 \begin{align*}
  L&={1\over 2}\sum_{i,j=1}^n\sum_{\mu\in S_i}\sum_{\nu\in S_j}D_{M+\nu,M+\mu}^\prime \left(-{1\over p_\mu^2}+{1\over p_\nu^2}\right)b_{\mu\nu}\\
                     &={1\over 8}\sum_{\substack{i,j=1\\ (i\not=j)}}^nc_{ij}\sum_{\mu\in S_i}\sum_{\nu\in S_j}D_{M+\nu,M+\mu}^\prime, \tag{3.43}
 \end{align*}
where in passing to the second line of (3.43), we used (3.14). It follows from (3.10b) and the formula (3.18) that
$$\sum_{\mu\in S_i}\sum_{\nu\in S_j}D_{M+\nu,M+\mu}^\prime=|D^\prime(-{\bf e}_i; {\bf e}_j)|=|D(-{\bf z},-{\bf e}_i;{\bf z}, {\bf e}_j)|, \eqno(3.44)$$
which, substituted in (3.43), gives
$$L={1\over 8}\sum_{\substack{i,j=1\\ (i\not=j)}}^nc_{ij}|D(-{\bf z},-{\bf e}_i;{\bf z}, {\bf e}_j)|. \eqno(3.45)$$
\par
On the other hand, using (3.38) and (3.39), the right-hand side of (3.40) reduces to
$$R\equiv \sum_{i,k=1}^MD_{ik}^\prime{a_{ik}\over p_i^2}+\sum_{i,k=1}^MD_{ik}{z_iz_k\over p_i^2}. \eqno(3.46)$$
We substitute the explicit form of $a_{ik}$ from (3.12) and take into account the symmetry $D_{ik}^\prime=D_{ki}^\prime$, the first term of $R$ is modified as
\begin{align*}
\sum_{i,k=1}^MD_{ik}^\prime{a_{ik}\over p_i^2}
&=-{1\over 2}\sum_{i,k=1}^MD_{ik}^\prime\left({1\over p_i^2}-{1\over p_k^2}\right){p_i-p_k\over p_i+p_k}z_iz_k \\
&={1\over 2}\sum_{i,k=1}^MD_{ik}^\prime\left({1\over p_i^2}-{2\over p_ip_k}+{1\over p_k^2}\right)z_iz_k. \tag{3.47} 
\end{align*}
It turns out by applying the formula (3.18) to (3.47) that
\begin{align*}
\sum_{i,k=1}^MD_{ik}^\prime{a_{ik}\over p_i^2}
&={1\over 2}|D^\prime(-{\bf z}; {\bf z}_{\tau\tau})|-|D^\prime(-{\bf z}_\tau; {\bf z}_{\tau})|+{1\over 2}|D^\prime(-{\bf z}_{\tau\tau}; {\bf z})|\\
&={1\over 2}|D(-{\bf z},-{\bf z}; {\bf z},{\bf z}_{\tau\tau})|-|D(-{\bf z},-{\bf z}_\tau; {\bf z},{\bf z}_{\tau})|+{1\over 2}|D(-{\bf z},-{\bf z}_{\tau\tau}; {\bf z},{\bf z})\\
&=-|D(-{\bf z},-{\bf z}_\tau; {\bf z},{\bf z}_{\tau})|, \tag{3.48}
\end{align*}
where in passing to the last line, we used the fact that any determinant which contains two identical rows (or columns) is zero.
The similar procedure applied to the second term of $R$ yields
$$\sum_{i,k=1}^MD_{ik}{z_iz_k\over p_i^2}=|D(-{\bf z}; {\bf z}_{\tau\tau})|.\eqno(3.49)$$
Adding (3.48) and (3.49), we finally obtain
$$R=|D(-{\bf z}; {\bf z}_{\tau\tau})|-|D(-{\bf z},-{\bf z}_\tau; {\bf z},{\bf z}_{\tau})|. \eqno(3.50)$$
The desired relation (3.35) follows immediately from (3.40), (3.45) and (3.50), completing the proof of Eq. (3.7). \par
\bigskip
\noindent{\bf D. Remarks}\par
\noindent{\bf 1.} Let $C=(c_{ij})_{1\leq i,j\leq n}$ be a real symmetric matrix whose dinagonal elements are zero, i.e., $c_{ii}=0 (i=1, 2, ..., n)$,
and $P=(p_{ij})_{1\leq i,j\leq n}$ a regular matrix. Then, under  appropriate orthogonal transformation $u_i=\sum_{j=1}^np_{ij}u_j^\prime$, the
quadratic form (1.7b) can be recast to a canonical form
$$F=\sum_{i=1}^pu_i^{\prime 2}-\sum_{i=1}^qu_{p+i}^{\prime 2}, \ (p+q\leq n), \eqno(3.51)$$
where $p(q)$ is the number of positive (negative) eigenvalues of $C$, and $p$ and $q$ are determined uniquely by $C$.$^{15}$ Note that
since ${\rm Tr}\,C=0$, $p\not=0$ and $q\not=0$. Under the same transformation, the system of bilinear equations (3.6) and (3.7) can be converted into the system
$$D_yD_\tau f\cdot g_i^\prime=fg_i^\prime, \ (i= 1, 2, ..., p+q), \eqno(3.52)$$
$$D_\tau^2f\cdot f={1\over 2}\left(\sum_{i=1}^pg_i^{\prime 2}-\sum_{i=1}^qg_{p+i}^{\prime 2}\right), \eqno(3.53)$$
where $u_i^\prime=g_i^\prime/f (i=1, 2, ..., p+q)$. For example, if $c_{ij}=1\ (i\not=j), c_{ii}=0$, then $p=1$ and $q=n-1$ since the eigenvalues of $C$ are $n-1$\,(simple root) and $-1$\,($(n-1)$-ple root).\par
\medskip
\noindent{\bf 2.} When $F$ is a positive definite quadratic form of $u_i\ (i=1, 2, ..., n)$, we can put $p=n$ and $q=0$ in (3.52) and (3.53) provided that $C$ has $n$ distinct
positive eigenvalues. 
The system corresponding to (1.7) becomes
$$u_{i,xt}=u_i+{1\over 2}\left[\left(\sum_{j=1}^nu_j^2\right)u_{i,x}\right]_x
, \ (i=1, 2, ..., n). \eqno(3.54)$$
If we consider the continuum limit $n\rightarrow \infty$ for (3.54), then we have a $(2+1)$-dimensional
nonlocal PDE of the form
$$u_{xt}=u+{1\over 2}\left(u_x\int_{-\infty}^\infty u^2dz\right)_x,\quad u=u(x,z,t). \eqno(3.55)$$
This equation is an analog of the $(2+1)$-dimensional nonlocal nonlinear Schr\"odinger equation
$${\rm i}u_t=u_{xx}+2u\int_{-\infty}^\infty |u|^2dz, \quad u=u(x,z,t), \eqno(3.56)$$
arising from a continuum limit of the multi-component nonlinear Schr\"odinger equation.$^{16,17}$
By means of the hodograph transformation
$$dy=rdx+\left(\int_{-\infty}^\infty u^2dz\right)\,rd\tau,\quad dt=d\tau, \eqno(3.57)$$
we obtain the parametric representation of the solution for Eq. (3.55)
$$u={g\over f},\quad x=y-2\,{f_\tau\over f}, \eqno(3.58)$$
where $f=f(y,\tau)$ and $g=g(y,z,\tau)$ satisfy the system of bilinear equations
$$D_yD_\tau f\cdot g=fg,\quad D_\tau^2f\cdot f={1\over 2}\int_{-\infty}^\infty g^2dz. \eqno(3.59)$$
We will discuss the integrability of the  equation (3.55) in a separate paper. \par
\medskip
\noindent{\bf 3.} The bilinear equation (3.7) takes the same form as that of a coupled modified 
Koreweg-de Vries equations proposed in Ref. 18  where the proof of the multisoliton solution has been
performed by lengthy calculations using various formulas of pfaffians. Here, we have provided  a novel
proof relying only on an elementary theory of determinants. \par
\medskip
\noindent{\bf 4.} The coupled PDEs proposed recently in Ref. 19
$$u_{i,xt}=u_i-\sum_{1\leq j<k\leq n}c_{jk}(u_{j,x}u_k-u_ju_{k,x})u_i,\ (i=1, 2, ..., n), \eqno(3.60)$$
where the coupling constants $c_{jk}$ are skew-symmetric, are transformed to the following system of
bilinear equations through the dependent variable transformations $u_i=g_i/f\ (i=1, 2, ..., n)$
$$D_xD_tf\cdot g_i=fg_i, \ (i=1, 2, ..., n), \eqno(3.61)$$
$$D_xD_tf\cdot f=\sum_{1\leq j<k\leq n}c_{jk}D_xg_j\cdot g_k. \eqno(3.62)$$
Recall that the bilinear equation (3.61) coincides with (3.6) if we replace the variables $x$ and $t$ by $y$ and $\tau$, respectively.
We conjecture that the multisoliton solution  of Eqs. (3.61) and (3.62) will be given by (3.16) where the matrix $B_M$ has the form
$$b_{\mu\nu}=-c_{ij}\,{p_\mu p_\nu\over p_\mu+p_\nu},\ \mu\in S_i, \nu\in S_j\ (\mu, \nu=1, 2, ..., M\ (\mu\not=\nu);
i, j=1, 2, ..., n\ (i\not=j)), \eqno(3.63)$$
in place of (3.14). Obviously, the corresponding tau-functions $f$ and $g_i$ satisfy Eq. (3.61) since its proof does not
depend on the explicit form of $B_M$ except that it is a skew-symmetric matrix with the constant elements. For the 2-component system, we have checked 
that  Eqs. (3.61) and (3.62) exhibit
the  2- and 3-soliton solutions, i.e.,  $M_1=M_2=2, M_1=M_2=3$.
The proof of the  general multisoliton solution will be reported elsewhere. \par
\bigskip
\noindent{\bf IV. TWO-COMPONENT SYSTEM} \par
Here, we consider the two-component system (1.8) in  detail. 
 We first  show the integrability of the system by constructing a Lax pair and then  present the multisoliton solution. 
We also discuss an integrable system (1.9) which is closely related to the system (1.8). \par
\noindent {\bf A. Integrability} \par
For the system (1.8), the equations (3.2) and (3.3) corresponding to Eqs. (1.8) read
$$x_{y\tau}=-{1\over 2}(uv)_y, \quad u_{y\tau}=x_yu,\quad v_{y\tau}=x_yv, \eqno(4.1)$$
where the first of these equations comes from the $y$-derivative of the second equation of (3.2) with $F=uv$.
The system of equations (4.1) can be derived from the compatibility condition of the system of linear PDEs
$${\Psi}_y=U{\Psi},\ {\Psi}_\tau=V{\Psi} \eqno(4.2a)$$
with
$$U=\lambda\begin{pmatrix}x_y&u_y\\
                                    v_y&-x_y\end{pmatrix},
                                     \quad V={1\over 2}\begin{pmatrix}0&-u\\
                                    v&0\end{pmatrix}
                                    +{1\over 4\lambda}\begin{pmatrix}1&0\\
                                    0&-1\end{pmatrix}, \eqno(4.2b)$$
 where $\lambda$ is a spectral parameter. Note in this expression that $x_y=\sqrt{1-u_yv_y}$.
Indeed, it follows from the condition $\Psi_{y\tau}=\Psi_{\tau y}$ that 
$$U_\tau-V_y+UV-VU=O, \eqno(4.3)$$
which yields Eqs. (4.1). Using (2.1b), we can rewrite (4.2) in terms of the original variables $x$ and $t$
$${\Psi}_x=\tilde U{\Psi},\ {\Psi}_t=\tilde V{\Psi}, \eqno(4.4a)$$
with
$$\tilde U=\lambda\begin{pmatrix}1&u_x\\
                                    v_x&-1\end{pmatrix},
                                     \quad \tilde V={1\over 2}\begin{pmatrix}0&-u\\
                                    v&0\end{pmatrix}
                                    +{1\over 4\lambda}\begin{pmatrix}1&0\\
                                    0&-1\end{pmatrix}
                                    +{\lambda\over 2}\begin{pmatrix}uv&uvu_x\\
                                    uvv_x&-uv\end{pmatrix}. \eqno(4.4b)$$
This  is a Lax pair for the system of equations (1.8). Note that when $u=v$, (4.4) reduces to the Lax pair
for the SP equation.$^{4}$ 
One can apply the inverse scattering transform (IST) method to establish the complete integrability of the system (1.8).
\par
\bigskip
\noindent{\bf B. Multisoliton solution}\par
\noindent {\bf 1. $N$-soliton solution} \par
The parametric representation of the multisoliton solution of Eqs. (1.8) is given by (3.5) with $n=2$
$$u={g_1\over f},\quad v={g_2\over f},\quad x=y-2\,{f_\tau\over f}. \eqno(4.5)$$
Here, we consider the case where both $u$ and $v$ contain $N$ solitons. Correspondingly, we set $M_1=M_2=N$ and $M=2N$ in (4.5).
The tau-functions $f$ and $g_i (i=1, 2)$ from (3.16) are represented by the following formulas:
$$f=\sqrt{F},\quad g_i=\sqrt{G_i}\ (i=1, 2), \eqno(4.6a)$$
with
$$F=|D|=\begin{vmatrix} A_{2N} & I_{2N} \\ -I_{2N} & B_{2N} \end{vmatrix}, \eqno(4.6b)$$
$$G_i=|D(-{\bf z}, -{\bf e}_i; {\bf z}, {\bf e}_i)|=\begin{vmatrix} A_{2N} & I_{2N} & {\bf z}^T & {\bf 0}^T \\
                                                       -I_{2N} & B_{2N} & {\bf 0}^T & {\bf e}_i^T \\
                                                        -{\bf z} & {\bf 0} &  0 & 0 \\
                                                        {\bf 0} & -{\bf e}_i &  0 & 0 \end{vmatrix},\ (i=1, 2). \eqno(4.6c)$$
 Here, the $2N\times 2N$ skew-symmetric matrices $A_{2N}$ and $B_{2N}$ have the elements
 $$A_{2N}=(a_{ij})_{1\leq i,j\leq 2N}, \quad a_{ij}=-{p_i-p_j\over p_i+p_j}\,{\rm e}^{\xi_i+\xi_j}\equiv -{p_i-p_j\over p_i+p_j}z_iz_j,$$
 $$\quad \xi_i=p_iy+{1\over p_i}\tau+\xi_{i0},\ (i=1, 2, ..., 2N), \eqno(4.6d)$$
$$B_{2N}=\begin{pmatrix} O_{N\times N} & B_{N\times N} \\ -B^T_{N\times N} & O_{N\times N}\end{pmatrix},$$
$$\quad B_{N\times N}=(b_{i\,N+j})_{1\leq i,j\leq N},\quad b_{i\,N+j}={1\over 4}{(p_ip_{N+j})^2\over p_i^2-p_{N+j}^2},\ (i, j=1, 2, ..., N), \eqno(4.6e)$$
and the $2N$-component vectors $\bf z$ and ${\bf
 e}_i\ (i=1, 2)$ are given respectively by
$${\bf z}=({\rm e}^{\xi_1}, {\rm e}^{\xi_2},..., {\rm e}^{\xi_{2N}}), \quad
{\bf e}_1=(\underbrace{1, 1,..., 1}_{N},\underbrace{0, 0,..., 0}_{N}), \quad {\bf e}_2=(\underbrace{0, 0,..., 0}_{N},\underbrace{1, 1,..., 1}_{N}). \eqno(4.6f)$$
Note that the $N$-soliton solution contains $4N$ complex-valued parameters $p_i,\ \xi_{i0}\ (i=1, 2, ..., 2N)$. An alternative parametrization 
with the same number of the parameters is
possible if one puts $p_{N+i}=p_i\ (i=1, 2, ..., N)$ and replace $\xi_{i0}$ and $\xi_{N+i\,0}$ by
$\xi_{i0}+{\rm ln}\,a_i$ and $\xi_{i\,0}+{\rm ln}\,b_i\ (i=1, 2, ..., N)$, respectively where $a_i$ and $b_i$ are new parameters. 
In the following, we present a few examples of solutions and investigate their properties. \par
\noindent {\bf 2. 1-loop soliton solution}\par
We give two types of tau-functions described above:
$$f=1+{1\over 4}{(p_1p_2)^2\over (p_1+p_2)^2}\,z_1z_2,\quad g_1=z_1,\quad g_2=z_2, \eqno(4.7a)$$
$$f=1+{a_1b_1p_1^2\over 16}z_1^2,\quad g_1=a_1z_1,\quad g_2=b_1z_1,\ (a_1, a_2, p_1 >0). \eqno(4.7b)$$
The solution corresponding to (4.7b) is calculated from (3.5) to give
$$u={2\over p_1}\sqrt{a_1\over b_1}\,{\rm sech}(\xi_1+\delta_1), \quad v={2\over p_1}\sqrt{b_1\over a_1}\,{\rm sech}(\xi_1+\delta_1), \eqno(4.8a)$$
$$x=y-{2\over p_1}\,{\tanh}(\xi_1+\delta_1),\quad \delta_1={\rm ln}\left({\sqrt{a_1b_1}p_1\over 4}\right), \ (a_1, b_1, p_1 >0). \eqno(4.8b)$$
A profile of $u$ is depicted in Fig. 1. It represents a loop soliton with the amplitude ${2\over p_1}\sqrt{a_1\over b_1}$ 
and the velocity $c_1=1/ p_1^2$. 
Note that the amplitude of the loop soliton is defined by the  maximum value of $u$ which is attained at $\xi_1=-\delta_1$ in the present example.
The property of $v$ is the same as that of $u$ except the amplitude given by ${2\over p_1}\sqrt{b_1\over a_1}$.
By comparing (2.16) and (4.8), we see that the loop soliton has the same structure as that of the loop soliton of the SP equation. \par
\begin{center}
\includegraphics[width=8cm]{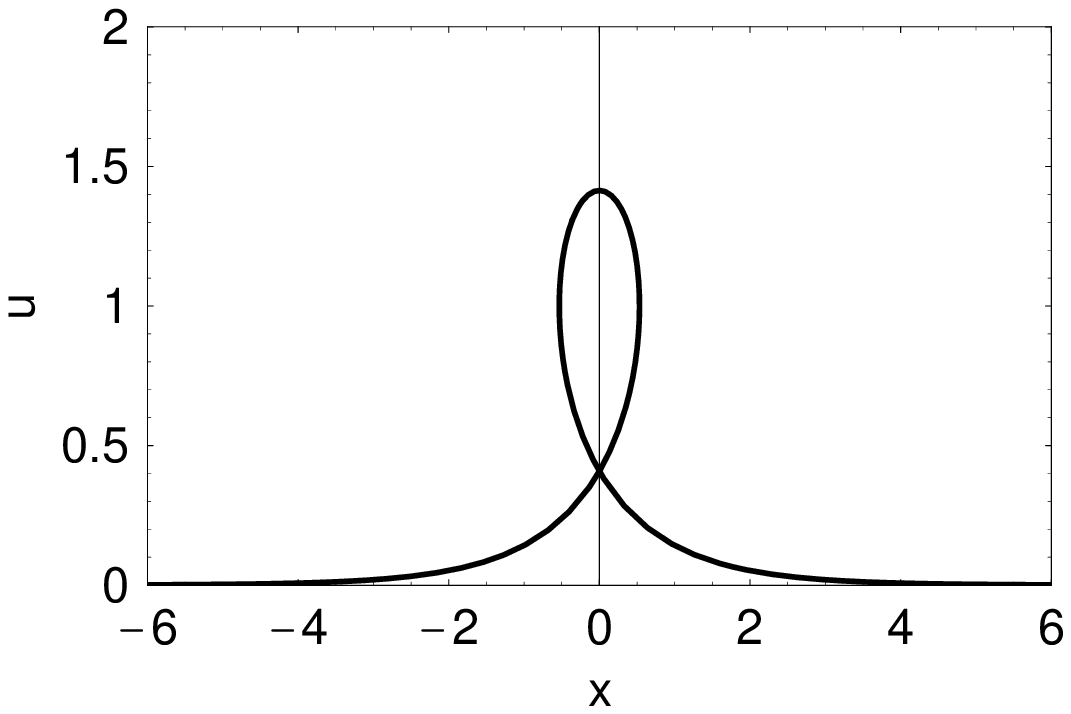}
\end{center}
{\bf FIG. 1.}\ The profile of the 1-loop soliton solution $u$  with the parameters $p_1=1.0, a_1=0.5$ and $b_1=1.0$. \par
\medskip
\noindent {\bf 3. 2-loop soliton solution}\par
As in the case of the 1-soliton solution, we write down two types of  tau-functions for the 2-soliton solution: \par
$$f=1+{1\over 4}{(p_1p_3)^2\over (p_1+p_3)^2}\,z_1z_3+ {1\over 4}{(p_1p_4)^2\over (p_1+p_4)^2}\,z_1z_4+{1\over 4}{(p_2p_3)^2\over (p_2+p_3)^2}\,z_2z_3
+{1\over 4}{(p_2p_4)^2\over (p_2+p_4)^2}\,z_2z_4$$
$$+{1\over 16}{(p_1p_2p_3p_4)^2(p_1-p_2)^2(p_3-p_4)^2\over (p_1+p_3)^2(p_2+p_3)^2(p_1+p_4)^2(p_2+p_4)^2}\,z_1z_2z_3z_4, \eqno(4.9a)$$
$$g_1=z_1+z_2+{1\over 4}{p_3^2(p_1-p_2)^2\over (p_1+p_3)^2(p_2+p_3)^2}\,z_1z_2z_3+{1\over 4}{p_4^4(p_1-p_2)^2\over (p_1+p_4)^2(p_2+p_4)^2}\,z_1z_2z_4, \eqno(4.9b)$$
$$g_2=z_3+z_4+{1\over 4}{p_1^4(p_3-p_4)^2\over (p_1+p_3)^2(p_1+p_4)^2}\,z_1z_3z_4+{1\over 4}{p_2^4(p_3-p_4)^2\over (p_2+p_3)^2(p_2+p_4)^2}\,z_2z_3z_4. \eqno(4.9c)$$
$$f=1+{1\over 16}a_1b_1p_1^2z_1^2+{1\over 4}(a_1b_2+a_2b_1){(p_1p_2)^2\over (p_1+p_2)^2}\,z_1z_2+{1\over 16}a_2b_2p_2^2z_2^2$$
$$+{1\over 256}a_1a_2b_1b_2{(p_1p_2)^2(p_1-p_2)^4\over (p_1+p_2)^4}\,(z_1z_2)^2, \eqno(4.10a)$$
$$g_1=a_1z_1+a_2z_2+{1\over 16}a_1a_2b_1{p_1^2(p_1-p_2)^2\over (p_1+p_2)^2}\,z_1^2z_2+{1\over 16}a_1a_2b_2{p_2^2(p_1-p_2)^2\over (p_1+p_2)^2}\,z_1z_2^2, \eqno(4.10b)$$
$$g_2=b_1z_1+b_2z_2+{1\over 16}a_1b_1b_2{p_1^2(p_1-p_2)^2\over (p_1+p_2)^2}\,z_1^2z_2+{1\over 16}a_2b_1b_2{p_2^2(p_1-p_2)^2\over (p_1+p_2)^2}\,z_1z_2^2. \eqno(4.10c)$$
We consider the 2-soliton solution corresponding to the tau-functions (4.10). Figure 2 shows the time evolution of of the 2-soliton  solution $u$. 
It represents the interaction of two loop solitons,
each takes the form of the 1-loop soliton given by (4.8), as we demonstrate now.
 \par
\begin{center}
\includegraphics[width=8cm]{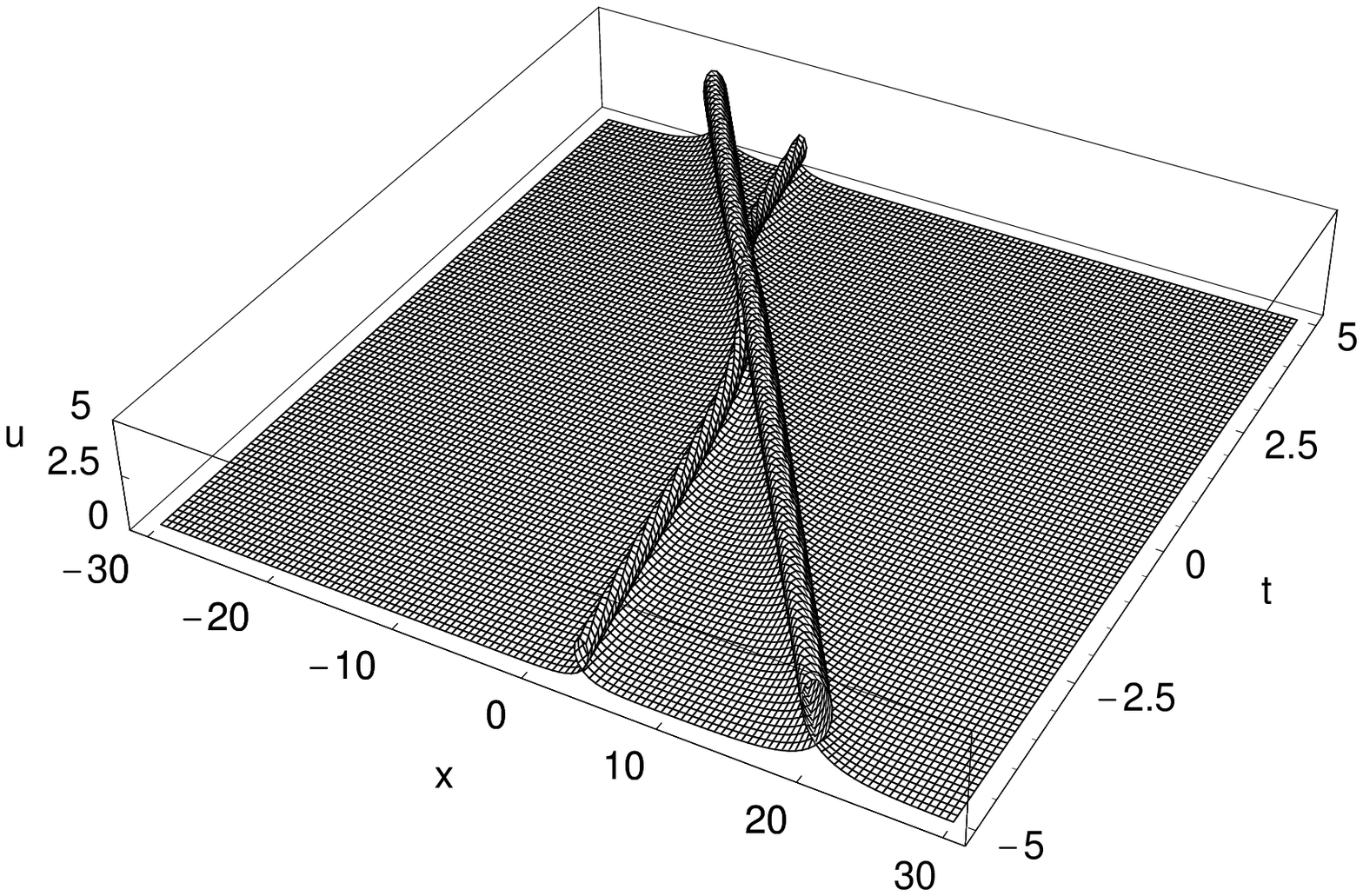}
\end{center}
{\bf FIG. 2.}\ The time evolution of the 2-loop soliton solution $u$ with the parameters $p_1=0.5, p_2=1.0, a_1=1.0, a_2=2.0, b_1=1.0, b_2=2.0$ and $x_{10}=x_{20}=0$. \par
\medskip
We investigate the asymptotic behavior of the solution $u$. To this end, we assume $0<p_1<p_2$ and $a_i>0, b_i>0\ (i=1, 2)$.  Then, an asymptotic analysis similar to
that developed for the 2-loop soliton solution of the SP equation shows that as $t\rightarrow -\infty$, $u$ behaves like$^{9}$
$$u=u_1+u_2 \sim {2\over p_1}\sqrt{a_1\over b_1}\, {\rm sech}(\xi_1+\delta_1^\prime)+{2\over p_2}\sqrt{a_2\over b_2}\, {\rm sech}(\xi_2+\delta_2), \eqno(4.11a)$$
$$x+c_1t-x_{10} \sim {\xi_1\over p_1}-{2\over p_1}\,{\rm tanh}(\xi_1+\delta_1^\prime)-{2\over p_1}-{4\over p_2}, \ {\rm for}\ u_1, \eqno(4.11b)$$
$$x+c_2t-x_{20} \sim {\xi_1\over p_2}-{2\over p_2}\,{\rm tanh}(\xi_2+\delta_2)-{2\over p_2}, \ {\rm for}\ u_2, \eqno(4.11c)$$
where
$$c_i={1\over p_i^2}, \quad \delta_i={\rm ln}\left({\sqrt{a_ib_i}\over 4}p_i\right), \quad \delta_i^\prime={\rm ln}\left[{\sqrt{a_ib_i}\over 4}p_i\left({p_1-p_2\over p_1+p_2}\right)^2\right],
    \ (i=1, 2). \eqno(4.11d)$$
As $t\rightarrow +\infty$, on the other hand
$$u=u_1+u_2 \sim {2\over p_1}\sqrt{a_1\over b_1}\, {\rm sech}(\xi_1+\delta_1)+{2\over p_2}\sqrt{a_2\over b_2}\, {\rm sech}(\xi_2+\delta_2^\prime), \eqno(4.12a)$$
$$x+c_1t-x_{10} \sim {\xi_1\over p_1}-{2\over p_1}\,{\rm tanh}(\xi_1+\delta_1)-{2\over p_1}, \ {\rm for}\ u_1, \eqno(4.12b)$$
$$x+c_2t-x_{20} \sim {\xi_1\over p_2}-{2\over p_2}\,{\rm tanh}(\xi_2+\delta_2^\prime)-{2\over p_2}-{4\over p_1}, \ {\rm for}\ u_2. \eqno(4.12c)$$
We observe that the solution $u$ splits into two loop solitons as time evolves, each of which has the form of single loop soliton.
The only effect due to the interaction of two loop solitons is the phase shift. To see this,
let $x_{ic}$ be the center position of the $i$th soliton. Then, it follows from the asymptotic forms (4.11) and (4.12) that
$$x_{1c}+c_1t-x_{10} \sim -{\delta_1^\prime \over p_1}-{2\over p_1}-{4\over p_2}, \quad x_{2c}+c_2t-x_{20} \sim -{\delta_2\over p_2}-{2\over p_2}, \ (t\rightarrow \ -\infty), \eqno(4.13a)$$
$$x_{1c}+c_1t-x_{10} \sim -{\delta_1\over p_1}-{2\over p_1}, \quad x_{2c}+c_2t-x_{20} \sim -{\delta_2^\prime\over p_2}-{2\over p_2}-{4\over p_1}. \ (t\rightarrow \ +\infty). \eqno(4.13b)$$
Since two solitons propagate to the left, the phase shift of the $i$th soliton can be defined by the relation
$$\Delta_i=x_{ic}(t\rightarrow  -\infty)-x_{ic}(t\rightarrow  +\infty), \ (i=1, 2). \eqno(4.14)$$
Thus, from (4.13) and (4.14) one has
$$\Delta_1=-{1\over p_1}\,{\rm ln}\left({p_1-p_2\over p_1+p_2}\right)^2 -{4\over p_2}, \eqno(4.15a)$$
$$\Delta_2={1\over p_2}\,{\rm ln}\left({p_1-p_2\over p_1+p_2}\right)^2 +{4\over p_1}. \eqno(4.15b)$$
The same calculation can be applied to $v$ as well. 
The corresponding asymptotic formulas are obtained simply by interchanging $a_i$ and $b_i\ (i=1, 2)$ in the above  expressions.
It should be remarked that the above formulas for the phase shift do not depend on $a_i$ and $b_i\ (i=1, 2)$ and are determined only by the amplitude
parameters $p_1$ and $p_2$. They coincide with  those of the 2-loop soliton solution of the SP equation.$^{9}$
A novel feature of the solution in the present 2-component system is that the large soliton propagates slower than the small soliton
if the inequality ${1\over p_1}\sqrt{a_1\over b_1}<{1\over p_2}\sqrt{a_2\over b_2}$ holds. This fact is seen from the asymptotic forms
(4.11a) and (4.12a) of the solution with the velocities of $u_1$ and $u_2$ being given respectively by $c_1=1/p_1^2$ and $c_2=1/p_2^2$\ $(c_2<c_1)$. \par
\newpage
\noindent{\bf 4. Breather solutions}\par
The breather solutions are constructed from the soliton solutions by following the same manipulation as that used for the 
soliton solutions of the SP equation.$^{9}$ 
Here, we present the 1-breather solution. In this case, we put
$$p_1=a+{\rm i}b=p_2^*,\ \xi_{10}=\lambda+{\rm i}\mu=\xi_{20}^*,
\  a_1=\alpha_1{\rm e}^{{\rm i}\phi_1}=a_2^*, \ b_1=\beta_1{\rm e}^{{\rm i}\psi_1}=b_2^*, \eqno(4.16)$$
in (4.10)  to obtain the tau-functions $f$, $g_1$ and $g_2$. 
Here,  $a, b, \alpha_1$ and $\beta_1$ are positive constants, $\lambda, \mu, \phi_1$ and $\psi_1$ are real constants and the asterisk denotes  complex conjugate.
After a few calculations, we find the following expressions:
$$f={4\over b^2\,}{\rm e}^{2(\theta+\theta_0)}\hat f, 
\quad g_1={16{a\over b}\sqrt{\alpha_1\over \beta_1}\over \sqrt{a^2+b^2}}\,{\rm e}^{2(\theta+\theta_0)}\,\hat g_1, 
\quad g_2={16{a\over b}\sqrt{\beta_1\over \alpha_1}\over \sqrt{a^2+b^2}}\,{\rm e}^{2(\theta+\theta_0)}\,\hat g_2, \eqno(4.17a)$$
with
$$\hat f=b^2\cosh^2(\theta+\theta_0)+a^2\cos^2(\chi+\chi_0+\delta^\prime)-(a^2+b^2)\,\sin^2\delta, \eqno(4.17b)$$
$$\hat g_1=\sin(\chi_0-\delta)\sin(\chi+\chi_0+\delta^\prime)\,\cosh(\theta+\theta_0)
-\,\cos(\chi_0-\delta)\cos(\chi+\chi_0+\delta^\prime)\sinh(\theta+\theta_0),\eqno(4.17c)$$
$$\hat g_2=\sin(\chi_0+\delta)\sin(\chi+\chi_0+\delta^\prime)\,\cosh(\theta+\theta_0)
-\,\cos(\chi_0+\delta)\cos(\chi+\chi_0+\delta^\prime)\sinh(\theta+\theta_0),\eqno(4.17d)$$
where
$$\theta=a\left(y+{1\over a^2+b^2}\,\tau\right)+\lambda,\quad \chi=b\left(y-{1\over a^2+b^2}\,\tau\right)+\mu, \eqno(4.18a)$$
$${\rm e}^{\theta_0}={b\over 4a}\sqrt{\alpha_1\beta_1(a^2+b^2)}, \quad \tan\,\chi_0={b\over a}, \quad \delta={1\over 2}(\phi_1-\psi_1),
\quad \delta^\prime={1\over 2}(\phi_1+\psi_1).  \eqno(4.18b)$$
 Substituting (4.17) into (3.5), we obtain the parametric
representation of the 1-breather solution:
$$u={4ab\sqrt{\alpha_1\over \beta_1}\over \sqrt{a^2+b^2}}\,{\hat g_1\over \hat f},
\quad v={4ab\sqrt{\beta_1\over \alpha_1}\over \sqrt{a^2+b^2}}\,{\hat g_2\over \hat f}, \eqno(4.19a)$$
$$x=y-{2ab\over a^2+b^2}\,{b\,\sinh\,2(\theta+\theta_0)+a\,\sin\,2(\chi+\chi_0+\delta^\prime)
\over \hat f}-{4a\over a^2+b^2}. \eqno(4.19b)$$
Both $u$ and $v$ include two different phases $\theta$ and $\chi$. The former characterizes the envelope of the breather whereas the latter
governs the internal oscillation. 
Figure 3 shows the time evolution of the 1-breather solution $u$. 
It represents a oscillating localized pulse moving to the left. Contrary to single loop soliton, the profile
of the pulse is nonstationary in the comoving coordinate system.
An inspection shows that the  solution (4.19)  exhibits singularities as encountered  in the case of the
breather solution of the SP equation.
 Therefore, certain condition must be imposed on the parameters $a$, $b$ and $\delta$ to produce the regular breather.
 However, we do not pursue the detail here.\par
\begin{center}
\includegraphics[width=8cm]{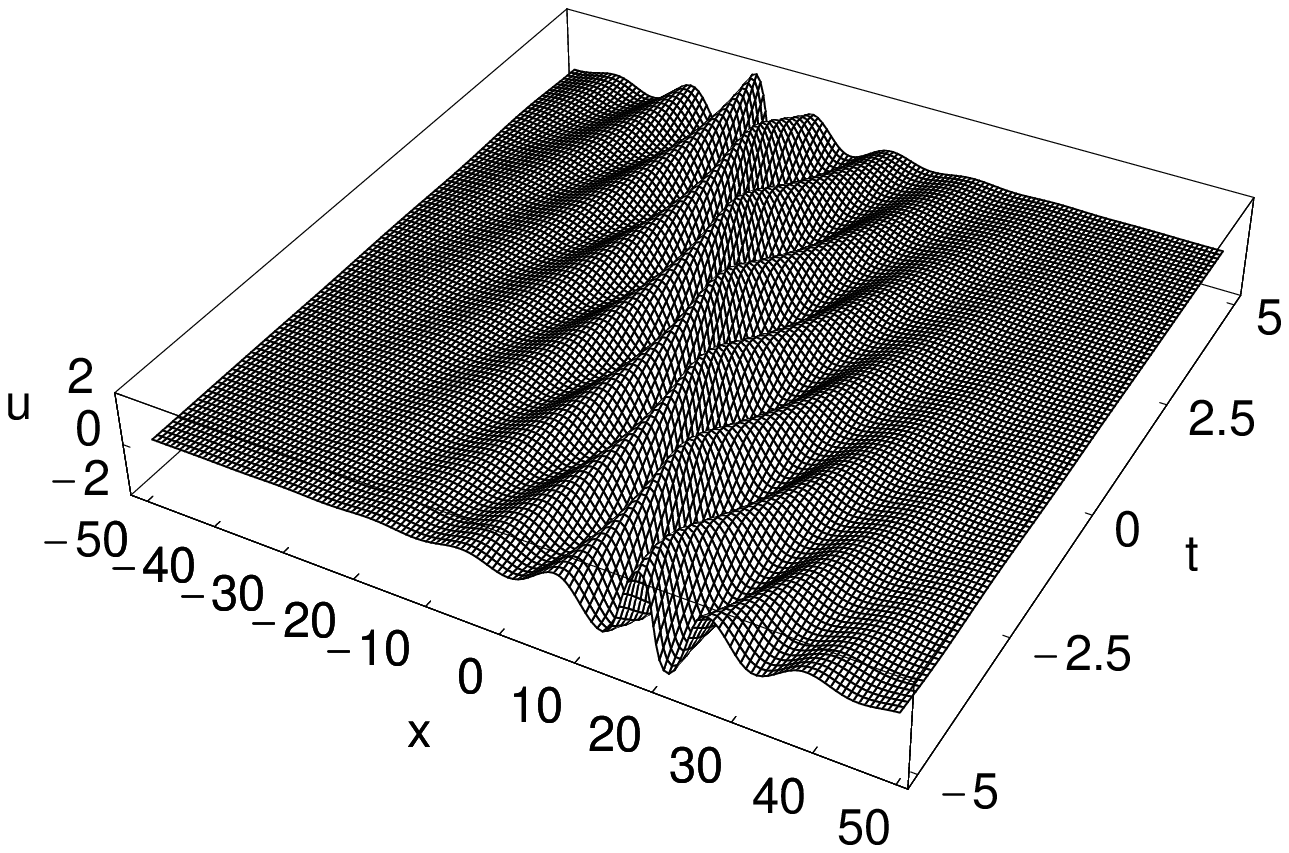}
\end{center}
{\bf FIG. 3.}\ The time evolution of the 1-breather solution $u$ with the parameters $a=0.1, b=0.5, \alpha_1=1.0, \beta_1=2.0, \phi_1=0, \psi_1=\pi/2,$
 and $\lambda=\mu=0$. \par
\medskip
\noindent {\bf C. Related integrable system}\par
\noindent{\bf 1. 2-component system}\par
The 2-component system (1.8) can be transformed to another integrable system (1.9) by a simple transformation. To show this, we put
$$u=\tilde u+{\rm i}\tilde v, \quad v=\tilde u-{\rm i}\tilde v, \eqno(4.20)$$
and substitute this into Eqs. (1.8),
 we obtain a system of equations for $\tilde u$ and $\tilde v$
$$\tilde u_{xt}=\tilde u+{1\over 2}[(\tilde u^2+\tilde v^2)\tilde u_x]_x, \eqno(4.21a)$$
$$\tilde v_{xt}=\tilde v+{1\over 2}[(\tilde u^2+\tilde v^2)\tilde v_x]_x. \eqno(4.21b)$$
This system is a special case of the $n$-component system (3.54) with $n=2$. \par
\noindent {\bf 2. $N$-soliton solution}\par
The parametric representation of the $N$-soliton solution for Eqs. (4.21) can be expressed in the form
$$\tilde u={\tilde g_1\over \tilde f},\quad \tilde v={\tilde g_2\over \tilde f},\quad x=y-2\,{\tilde f_\tau\over \tilde f}, \eqno(4.22)$$
where the tau-functions $\tilde g_1, \tilde g_2$ and $\tilde f$ satisfy the system of bilinear equations
$$D_yD_\tau\tilde f\cdot \tilde g_i = \tilde f\tilde g_i,\ (i=1, 2), \eqno(4.23a)$$
$$D_\tau^2\tilde f\cdot \tilde f={1\over 2}(\tilde g_1^2+\tilde g_2^2). \eqno(4.23b)$$
Here, we consider real-valued solutions $\tilde u$ and $\tilde v$ for the system (4.21).
The tau-functions representing the $N$-soliton solution are obtained 
from (4.6) by putting $p_{i+N}=p_i^*,\ \xi_{i+N\,0}=\xi_{i0}^*\ (i=1, 2, ..., N)$. Then, the expressions
corresponding to (4.6d), (4.6e) and (4.6f)  become
$$A_{2N}=\begin{pmatrix} A_1 & A_2\\ A_2^* & A_1^* \end{pmatrix},\quad A_1=\left(-{p_i-p_j\over p_i+p_j}z_iz_j\right)_{1\leq i,j\leq N},
\quad A_{2}=\left(-{p_i-p_j^*\over p_i+p_j^*\,}z_iz_j^*\right)_{1\leq i,j\leq N}, \eqno(4.24a)$$
$$B_{2N}=\begin{pmatrix} O_{N\times N} & B_1 \\ B_1^* & O_{N\times N}\end{pmatrix}, 
\quad  B_1=\left({1\over 4}\,{(p_ip_j^*)^2\over p_i^2-p_j^{*2}}\right)_{1\leq i,j\leq N}, \eqno(4.24b)$$
$${\bf z}=({\rm e}^{\xi_1}, {\rm e}^{\xi_2},..., {\rm e}^{\xi_{N}},{\rm e}^{\xi_1^*}, {\rm e}^{\xi_2^*},..., {\rm e}^{\xi_N^*}). \eqno(4.24c)$$
We can see that the tau-functions $f, g_1$ and $g_2$  (4.6) with (4.24) satisfy the conditions $f=f^*$ and $g_2=g_1^*$, which,
combined with (4.5) and (4.20), gives
$$\tilde f=f, \quad \tilde g_1={1\over 2}(g_1+g_1^*),\quad \tilde g_2={1\over 2{\rm i}}(g_1-g_1^*). \eqno(4.25)$$
\par
As in the case of the $N$-soliton solution of the 2-component system (1.8) (see Sec. B), we have an alternative parametrization of the
$N$-soliton solution. Namely, we replace $\xi_{j0}$ by $\xi_{j0}+ {\rm ln}(a_j+{\rm i}b_j)\ (j=1, 2, ..., N)$ 
where $a_j$ and $b_j$ are real parameters,
and then take $p_j$ and $\xi_{j0}\ (j=1, 2, ..., N)$
be real  in the expressions of the tau-functions. This procedure yields the tau-functions corresponding to (4.7b) and (4.10), for example.
Actually, for the 1-soliton solution, the corresponding tau-functions are given by
$$\tilde f=1+{1\over 16}(a_1^2+b_1^2)p_1^2z_1^2,\quad \tilde g_1=a_1z_1,\quad \tilde g_2=b_1z_1, \eqno(4.26)$$
and for the 2-soliton solution, they read
$$\tilde f=1+{1\over 16}(a_1^2+b_1^2)p_1^2z_1^2+{1\over 2}(a_1a_2+b_1b_2){(p_1p_2)^2\over (p_1+p_2)^2}\,z_1z_2+{1\over 16}(a_2^2+b_2^2)p_2^2z_2^2$$
$$+{1\over 256}(a_1^2+b_1^2)(a_2^2+b_2^2){(p_1p_2)^2(p_1-p_2)^4\over (p_1+p_2)^4}\,(z_1z_2)^2, \eqno(4.27a)$$
$$\tilde g_1=a_1z_1+a_2z_2+{1\over 16}a_2(a_1^2+b_1^2){p_1^2(p_1-p_2)^2\over (p_1+p_2)^2}\,z_1^2z_2+{1\over 16}a_1(a_2^2+b_2^2){p_2^2(p_1-p_2)^2\over (p_1+p_2)^2}\,z_1z_2^2, \eqno(4.27b)$$
$$\tilde g_2=b_1z_1+b_2z_2+{1\over 16}b_2(a_1^2+b_1^2){p_1^2(p_1-p_2)^2\over (p_1+p_2)^2}\,z_1^2z_2+{1\over 16}b_1(a_2^2+b_2^2){p_2^2(p_1-p_2)^2\over (p_1+p_2)^2}\,z_1z_2^2. \eqno(4.27c)$$
It can be seen that substitution of (4.26) and (4.27) into (4.22) produces the 1- and 2-loop soliton solutions, respectively. \par
\noindent{\bf 3. Breather solutions}\par
One can confirm by direct calculation that the  tau-functions (4.27) satisfy the bilinear equations (4.23).
 In the process, the reality of the parameters has not been used. This fact enables us to extend the range of the parameters to complex values. 
 Thus, the breather solutions are constructed
 from the soliton solutions by applying the procedure developed in Sec.IV  B. In practice, according to the
 parametrization (4.16), one has
 $$\tilde f={4\over b^2\,}{\rm e}^{2(\theta+\theta_0)}\bar f, 
\quad \tilde g_1={16\alpha_1a \over \gamma b\sqrt{a^2+b^2}}\,{\rm e}^{2(\theta+\theta_0)}\,\bar g_1, 
\quad \tilde g_2={16\beta_1a \over \gamma b\sqrt{a^2+b^2}}\,{\rm e}^{2(\theta+\theta_0)}\,\bar g_2, \eqno(4.28a)$$
with
$$\bar f=b^2\cosh^2(\theta+\theta_1)+a^2\cos^2(\chi+\chi_0+\kappa)+{1\over 2}(a^2+b^2)\left({\alpha_1^2+\beta_1^2\over \gamma^2}-1\right), \eqno(4.28b)$$
$$\bar g_1=\sin(\chi_0-\phi_1+\kappa)\sin(\chi+\chi_0+\kappa)\,\cosh(\theta+\theta_1)
-\,\cos(\chi_0-\phi_1+\kappa)\cos(\chi+\chi_0+\kappa)\sinh(\theta+\theta_1),\eqno(4.28c)$$
$$\bar g_2=\sin(\chi_0-\psi_1+\kappa)\sin(\chi+\chi_0+\kappa)\,\cosh(\theta+\theta_1)
-\,\cos(\chi_0-\psi_1+\kappa)\cos(\chi+\chi_0+\kappa)\sinh(\theta+\theta_1),\eqno(4.28d)$$
where the parameters $\theta_1, \gamma$ and $\kappa$ are defined by
$${\rm e}^{\theta_1}={b\over 4a}\sqrt{a^2+b^2}\,\gamma,\quad \gamma=[\alpha_1^4+2\alpha_1^2\beta_1^2\,\cos\,2(\phi_1-\psi_1)+\beta_1^4]^{1\over 4}, \eqno(4.28e)$$
$$\tan\,2\kappa={\alpha_1^2\,\sin\,2\phi_1+\beta_1^2\,\sin\,2\psi_1\over \alpha_1^2\,\cos\,2\phi_1+\beta_1^2\,\cos\,2\phi_1},\eqno(4.28f)$$
and $\theta, \chi$ and $\theta_0$ are already given respectively by (4.18a) and (4.18b).
 Substituting (4.28) into (4.22), we obtain the parametric
representation of the 1-breather solution:
$$\tilde u={4\alpha_1ab\over \gamma  \sqrt{a^2+b^2}}\,{\bar g_1\over \bar f},
\quad \tilde v={4\beta_1ab\over \gamma \sqrt{a^2+b^2}}\,{\bar g_2\over \bar f}, \eqno(4.29a)$$
$$x=y-{2ab\over a^2+b^2}\,{b\,\sinh\,2(\theta+\theta_1)+a\,\sin\,2(\chi+\chi_0+\kappa)
\over \bar f}-{4a\over a^2+b^2}. \eqno(4.29b)$$
\par
Of particular interest is a circularly polarized wave for which the solution exhibits a simple structure, as we shall now demonstrate.
  In this case, we put
 $\alpha_1=\beta_1, \chi_0-\phi_1+\kappa={\pi\over 2}$ and $\phi_1-\psi_1={\pi\over 2}$ to obtain the tau-functions
 $$\tilde f=1+{\alpha_1^2(a^2+b^2)^2\over 4a^2}\,{\rm e}^{2\theta},\quad \tilde g_1=2\alpha_1\,{\rm e}^\theta\cos(\chi+\phi_1), 
 \quad \tilde g_2=2\alpha_1\,{\rm e}^\theta\sin(\chi+\phi_1). \eqno(4.30)$$
 Then, the solution takes the form
 $$\tilde u={2a\over a^2+b^2}{\cos(\chi+\phi_1)\over \cosh(\theta+\theta_0^\prime)},
 \quad \tilde v={2a\over a^2+b^2}{\sin(\chi+\phi_1)\over \cosh(\theta+\theta_0^\prime)},\quad \left({\rm e}^{\theta_0^\prime}={\alpha_1(a^2+b^2)\over 2a}\right), \eqno(4.31a)$$
 $$x=y-{2a\over a^2+b^2}\,\tanh(\theta+\theta_0^\prime)-{2a\over a^2+b^2}. \eqno(4.31b)$$
 The parametric solution (4.31) represents a nonsingular breather if the inequality $0<a/b<1$ holds. 
  Figure 4 shows the time evolution of $u(\equiv \tilde u)$ given by (4.31). 
  \par
 \begin{center}
\includegraphics[width=8cm]{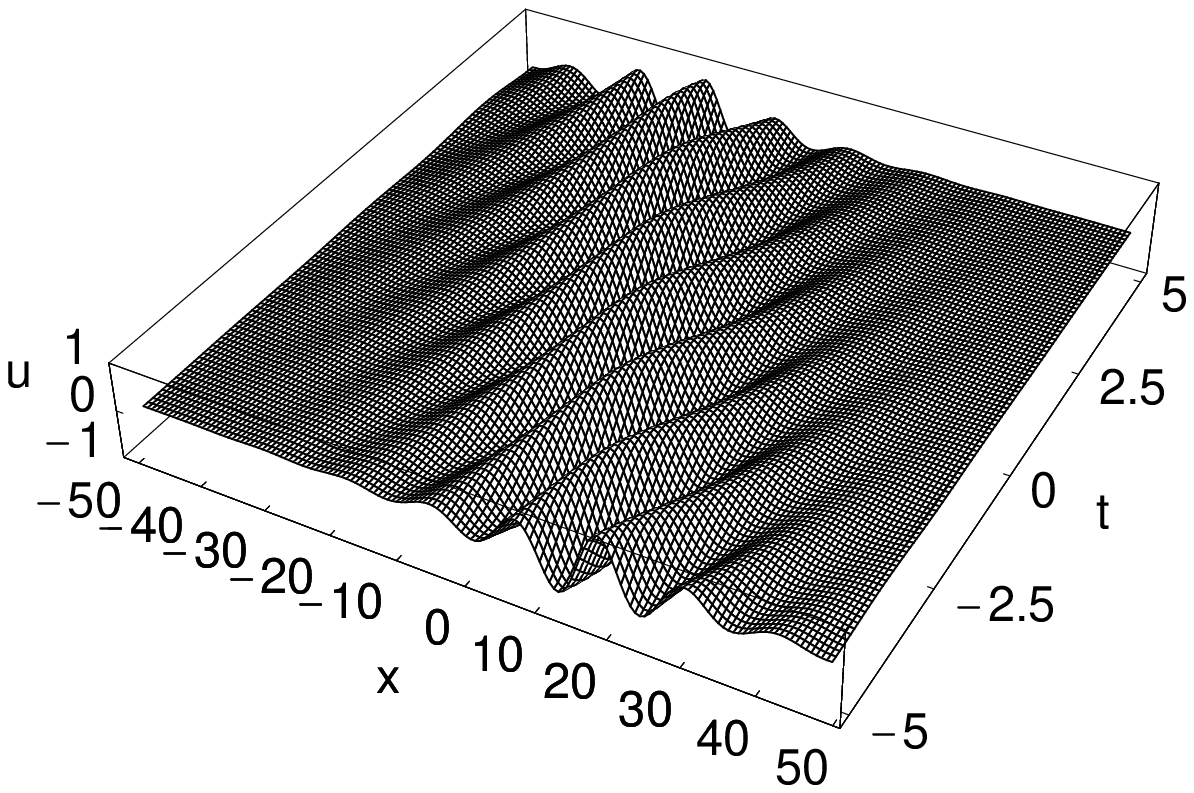}
\end{center}
{\bf FIG. 4.}\ The time evolution of the 1-breather solution $u(\equiv \tilde u)$ with the parameters $a=0.1, b=0.5, \alpha_1=1.0, \phi_1=0$ and $\lambda=\mu=0$. \par
 \medskip
  One can show by a direct calculation that the solution (4.31) satisfies  the  integral relations
 $$\int^\infty_{-\infty}\tilde u\,dx=0,\quad \int^\infty_{-\infty}\tilde v\,dx=0, \eqno(4.32)$$
 implying that both $\tilde u$ and $\tilde v$ are  zero mean fields. This fact
  indicates clearly an oscillating character of the solution.
 Note that the above relations represent the conservation laws derived from the system of equations (4.21) for localized waves.
It is interesting to observe that in the small amplitude limit $a/b\rightarrow 0$, the profile of $\tilde u$ bears resemblance to that of the soliton solution of the
nonlinear Schr\"odinger equation. \par
 \bigskip
 \noindent {\bf D. Remarks}\par
 \noindent {\bf 1.} The system of equations (4.1) is equivalent to a coupled dispersionless system for the
 variables $r=r(x,t)$, $s=s(x,t)$ and $q=q(x,t)$
 $$q_{xt}+(rs)_x=0,\quad r_{xt}-2q_xr=0,\quad s_{xt}-2q_xs=0. \eqno(4.33)$$
 The Lax pair associated with the system (4.33) has been obtained and the IST has been applied to it to
 construct soliton solutions.$^{20}$ In particular, 1- and 2-soliton solutions have been presented for $r$, $s$ and $q_x$. Here, we present
 the formula for the general multisoliton solution for the first time. \par
 \medskip
 \noindent {\bf 2.} The system of bilinear equations (4.23) can be derived from the system (4.33) with  a reduction $s=r^*$ 
 through appropriate dependent variable transformations.$^{21}$
See also an analysis by means of the IST.$^{22}$ \par
\bigskip
\noindent {\bf V. CONCLUSION} \par
In this paper, we proposed  a novel multi-component system associated with the SP equation and constructed its
multisoliton solutions in terms of pfaffians.  We also considered the equations reduced from our system. In particular,
 the 2-component system (1.8)  was found to be completely integrable 
for which  the explicit Lax pair was presented. We also provided the loop soliton and breather solutions for the system  and investigated their properties.
We also addressed the system (1.9) which stems from the system (1.8) by a simple transformation.
In conclusion, we shall discuss some open problems associated with the multi-component system under consideration. \par
\noindent {\bf 1.} One interesting issue to be resolved in a future work is the proof of the complete integrability of the
$n$-component system (1.7) by using the IST. To construct the Lax pair for the 
system, one way will be
to start from the system of bilinear equations (3.6) and (3.7) to obtain the B\"acklund transformation among the
tau-functions and then derive the scheme of the IST following the standard procedure in the bilinear formalism.  \par
\noindent {\bf 2.} Other issues to be reserved for detailed study have already been described in Sec. III\,D. Of particular importance
is the construction of the multisoliton solution of the $n$-component system (3.54) with $n\geq 3$. 
Unlike the 2-component system, the linear transformation such as (4.20) does not exist to convert the system (1.7) to the system (3.54).
Hence, one must solve the system of equations (3.52) and (3.53) with $p=n$ and $q=0$.  
It will be a relatively simple task to obtain the 1- and 2-soliton solutions analogous to (4.26) and (4.27). Nevertheless, a systematic
approach is necessary to construct general multisoliton solutions. \par
\noindent {\bf 3.} The system of equations (1.5) has been derived as a unidirectional model describing the propagation of circularly polarized ultra-short pulses
in a Kerr medium.$^{23}$ The solution of breather type has been obtained by means of an analysis as well as numerical computations.$^{24, 25}$
However, in view of the extremely 
complicated structure of the breather solution,$^{24}$ it seems to be unlikely that the system admits multibreather solutions as well. 
Thus, we suspect the complete integrability of the system even if it has passed the
Painlev\'e test. On the other hand,  although the difference between  (1.5) and (1.9) is the location of the 
$x$ deivative on the right-hand side,  the latter  shares many common features to the integrable systems such as the complete integrability and the exsistence of multisoliton
solutions. At present, however, the relevance of the system to the description of the dynamics of ultra-short pulses in optical fibers is not clear.
Nevertheless, it would be of interest to examine the possibility of the system as a physical model for the two-component generalization of the SP equation.
\par
\bigskip
\noindent {\bf ACKNOWLEDGEMENT}\par
This work was partially supported by the Grant-in-Aid for Scientific Research (C) No. 22540228 from Japan Society for the Promotion of Science. \par
\newpage
\noindent {\bf REFERENCES}\par
\begin{enumerate}[1.]
\item T. Sch\"afer and C. E. Wayne, Physica D {\bf 196}, 90 (2004).
\item  M. L. Rabelo, Stud. Appl. Math. {\bf 81}, 221 (1989).
\item R. Beals, M. Rabelo and K. Tenenblat, Stud. Appl. Math. {\bf 81}, 125 (1989).
\item A. Sakovich and S. Sakovich, J. Phys. Soc. Jpn. {\bf 74}, 239  (2005).
\item J. C. Brunelli, J. Math. Phys. {\bf 46}, 123507 (2005). \item J.C. Brunelli, Phys. Lett.  A {\bf 353}, 475 (2006).
\item A. Sakovich and S. Sakovich, J. Phys. A {\bf 39}, L361 (2006).
\item V. K. Kuetche, T. B. Bouetou and T. C. Kofane, J. Phys. Soc. Jpn. {\bf 76}, 024004 (2007).
\item Y. Matsuno, J. Phys. Soc. Jpn. {\bf 76}, 084003 (2007).
\item Y. Matsuno, J. Math. Phys. {\bf 49}, 073508 (2008).
\item Y. Matsuno, {\it Handbook of Solitons: Research, Technology and Applications}, edited by S. P. Lang and H. Bedore (Nova, New York, 2009) p. 541.
\item M. Pietrzyk, I. Kanatt\v{s}ikov and U. Bandelow, J. Nonl. Math. Phys. {\bf 15}, 162 (2008).
\item  S. Sakovich, J. Phys. Soc. Jpn. {\bf 77}, 123001 (2008)
\item R. Vein and P. Dale, {\it Determinants and Their Applications in Mathematical Physics} (Springer, New York, 1999)
\item I. Satake, {\it Linear Algebra} (Dekker, New York, 1975)
\item V. E. Zakharov, {\it Solitons}, Topics in Current Physics No. 17, edited by R. K. Bullough and P. J. Caudrey (Springer, Berlin, 1980) p. 243.
\item K. Maruno and Y. Ohta, Phys. Lett. A {\bf 372}, 4446 (2008).
\item M. Iwao and R. Hirota, J. Phys. Soc. Jpn. {\bf 66}, 577 (1997).
\item T. Tsuchida, J. Math. Phys. {\bf 51}, 053511 (2010).
\item K. Konno and H. Kakuhata, J. Phys. Soc. Jpn. {\bf 65}, 713 (1996).
\item H. Kakuhata and K. Konno, Theor. Math. Phys. {\bf 133}, 1675 (2002).
\item K. Konno, Appl. Anal. {\bf 57}, 209 (1993).
\item S.A. Kozlov and S.V. Sazonov, JETP {\bf 84}, 221 (1997).
\item D.V. Kartashov, A.V. Kim and S.A. Skobelev, JETP Lett. {\bf 78}, 276 (2003).
\item S.A. Skobelev and A.V. Kim, JETP Lett. {\bf 80}, 623 (2004).
\end{enumerate}

\end{document}